\documentclass{article}

\usepackage{amssymb}

\begin{document}

      \date{\today}

\begin{titlepage}

\title{Mining Metrics for Buried Treasure}

\author{D A Konkowski\thanks{dak@usna.edu}\\ Department of 
Mathematics, \\ U.S. Naval Academy, \\ Annapolis, Maryland, 21402 U.S.A. 
\and T M  Helliwell\thanks{T\_Helliwell@HMC.edu}\\ Department of Physics, 
\\ Harvey Mudd College, \\ Claremont, California 91711 USA}

\maketitle

\begin{abstract}

        The same but different: That might describe two metrics. On the 
surface CLASSI may show two metrics are locally equivalent, but buried 
beneath may be a wealth of further structure. This was beautifully described 
in a paper by M.A.H. MacCallum in 1998. Here I will illustrate the effect with 
two flat metrics -- one describing ordinary Minkowski spacetime and the 
other describing a three-parameter family of Gal'tsov-Letelier-Tod 
spacetimes. I will dig out the beautiful hidden classical singularity structure 
of the latter (a structure first noticed by Tod in 1994) and then show how 
quantum considerations can illuminate the riches. I will then discuss how 
quantum structure can help us understand classical singularities and metric 
parameters in a variety of exact solutions mined from the Exact Solutions 
book.

\end{abstract}


\maketitle

\end{titlepage}

 \section{Introduction}

	 \par I am very happy to be here to talk at Malcolm's 60th birthday 
celebration. This talk is a belated 60th birthday present. Happy Birthday, 
Malcolm!

	\par This talk will focus on cylindrically symmetric spacetimes 
\cite{SKMHH} whose metrics contain buried treasure: essential parameters 
that DO NOT appear in the Cartan scalars, i.e., the scalars that appear in the 
Cartan equivalence method for spacetime classification (see, e.g., 
\cite{MM1}). As the exact solutions book \cite{SKMHH} says, ``The 
method...due to Cartan...gives sets of scalars providing a unique {\it local} 
characterization, and thus leads to a procedure for comparing metrics." The 
emphasis on ``local" is mine; it is the key to the possibility of buried 
treasure. As Malcolm says in his 1998 paper \cite{MM} ``the metric may 
have parameters which are important globally but do not appear in the 
Cartan scalars" and ``The parameters cannot change the values of the 
Cartan scalars defined by the Riemann tensor and its derivatives at a point, 
and this directs attention to the possible global holonomy found by taking 
suitable closed curves..."

	\par The essential (buried) parameters:

\begin{itemize}

\item are unique to the characterization of the geometry, 

\item do not appear in the Cartan scalars (i.e., the spacetimes are locally 
equivalent), 

\item do appear in expressions of the (linear or affine) holonomy (i.e., the 
spacetimes are globally inequivalent), and \item may be due to a singular axis 
(requiring a relaxation of the usual definition of cylindrical symmetry -- see, 
e.g., Mars and Senovilla \cite{MS}, Carot et al \cite{Carot}, and SKMHH 
\cite{SKMHH}) or may be necessary to match to a regular source in the 
interior.

\end{itemize}

\par There is a great ``flat" example: the three-parameter family of 
Gal'tsov-Letelier-Tod (GLT) spacetimes \cite{GL, Tod},

 \begin{equation}      ds^{2} = -(dt + \alpha d\phi)^{2} + dr^{2} + 
\beta^{2}r^{2}d\phi^{2} + (dz + \gamma d\phi)^{2}.  \end{equation}

\noindent The coordinate ranges are the usual: $-\infty < t < \infty$, $0 \le 
r < \infty$, $0\le \phi \le 2\pi$, and $-\infty < z < \infty.$ Here $\alpha, 
\beta, \gamma$ are constants. They are ``the essential parameters". 
These spacetimes will be used as the key examples in the first half of this 
talk: a local analysis, as in CLASSI, does not distinguish these metrics from 
Minkowski spacetime although, as we shall see, there is a wealth of global 
structure hidden in this three-parameter family.

\par The plan of this talk is the following: After (1) the Introduction, (2) the 
classical structure of the GLT spacetimes will be described. (3) The global 
structure of spacetimes will be reviewed as is necessary to understand 
classical and quantum singularities. The quantum singularity of (4) GLT 
spacetimes and (5)  Special Cases will be discussed. (6) General cylindrically 
symmetric static spacetimes with disclinations and dislocations will be 
considered and followed by studies of (7) generalized Levi-Civita spacetimes 
with dislocations, (8) Chitre et al spacetimes, (9) the Melvin universes and 
(10) generalized Raychaudhuri spacetimes with disclinations and dislocations. 
(11) A final discussion will conclude the talk.

\section{GLT Spacetimes -- Classical Aspects}

\par Gal'tsov and Letelier in 1993 \cite{GL} and Tod in 1994 \cite{Tod} 
completely analyzed the spacetimes described in Eq.(1). As Tod noted: this 
three-parameter family of spacetimes describe a multitude of physically 
interesting cases. If $\alpha =0, \gamma = 0, \beta^{2} \neq 1$, they 
describe the idealized cosmic string \cite{FV, Gott, Hiscock, Linet}. If 
$\alpha =0$ and the final term is absent, they describe the ``point source" 
in $2+1$ gravity \cite{Clement1, Deser, Brown}. If $\alpha \neq 0$ and the 
final term is absent, they describe the ``rotating point source" of $2+1$ 
gravity \cite{Clement2, Brown}. And, if $\alpha = 0, \gamma \neq 0$, the 
GLT spacetime is the asymptotic metric \cite{BCM} at a large spatial 
separation from a cylindrically symmetric gravitational wave.

\par Here we will specialize our discussion to the static case ($\alpha = 0$) 
\cite{HK, HKA} where the metric takes the form,

\begin{equation}      ds^{2} = -dt^{2} + dr^{2} + \beta^{2}r^{2}d\phi^{2} + 
(dz + \gamma d\phi)^{2}. \end{equation}

\noindent This metric is classically singular if $\beta^{2} \neq 1$ and/or 
$\gamma \neq 0$; in these cases there is a quasiregular singularity at 
$r=0$. For clarity we will consider two special cases: an idealized cosmic 
string and a screw dislocation spacetime.

\par An idealized cosmic string is described if $\beta^{2} \neq 1$ and 
$\gamma = 0$ in Eq.(2). In this case there are incomplete geodesics which 
hit $r=0$ which is a quasiregular singularity, a disclination in crystallographic 
terminology (see, e.g., \cite{PS}). There is non-trivial linear holonomy and 
$r=0$ is a $\delta$-function in curvature \cite{Tod, Garfinkle}.

\par The second special case is the screw dislocation spacetime where 
$\gamma \neq 0$ and $\beta^{2} =1$ in Eq.(2). Incomplete geodesics hit 
$r=0$. A curve of bounded acceleration goes to $r=0, z=\infty$ in finite 
affine length. There is a quasiregular singularity at $r=0$ which is called a 
dislocation in crystallographic terminology. There is non-trivial affine 
holonomy and $r=0$ is a $\delta$-function in torsion \cite{Tod}.

\section{Global Structure -- Singularities}

\par In a maximal spacetime,

\begin{itemize}

\item A {\it classical singularity} exists if there are incomplete geodesics or 
incomplete paths of bounded acceleration \cite{ES, TCE, HE}.

\item A {\it quantum singularity} exists if the evolution of a test wave 
packet is not uniquely defined by the initial wave packet, without having to 
add information not present in the wave operator, spacetime metric and 
manifold alone (i.e., one must add boundary conditions at the singularity) 
\cite{HM, Wald, HK, HKA}.

\end{itemize}

\par Given the two categories of singularities various questions arise: (1) 
Are all classically singular spacetimes quantum mechanically singular as well? 
Answer: No. (see Horowitz and Marolf, 1995 \cite{HM}); (2) Since the 
topological parameters (e.g., $\alpha, \beta, \gamma$ in GLT spacetime) 
affect the existence of a classical singularity, do they affect the existence 
of a quantum singularity as well? We will consider this latter question after 
briefly discussing classical and quantum singularities in more detail.

\subsection{Classical Singularities}

\par In classical general relativity singularities are not part of the spacetime 
(the manifold is smooth): they are boundary points in an otherwise maximal 
spacetime \cite{ES}. For the timelike and null geodesics (or curves of 
bounded acceleration) that hit these boundary points, there is an 
incompleteness, an abrupt ending to the classical particle paths. The 
classical singularities which occur in otherwise maximal spacetimes have 
been divided into three types by Ellis and Schmidt \cite{ES}:

\begin{itemize}

\item {\it quasiregular} (e.g., the 2D cone, the idealized cosmic string)

\item {\it non-scalar curvature} (e.g. whimper cosmologies)

\item {\it scalar curvature} (e.g., the center of a Schwarzschild black hole, 
the beginning of a classical Big Bang cosmology.)

\end{itemize}

\par What if quantum wave packets are used instead of classical particles to 
test for a singularity? A quantum singularity would have to be described by 
ill-posed wave propogation. We'll see next how this has been defined.

\subsection{Quantum Singularities}

\par According to Horowitz and Marolf \cite{HM}, a static spacetime is 
quantum mechanically singular if the spatial portion of the Klein-Gordon wave 
operator is not essentially self-adjoint \cite{RS} on a $C_{0}^{\infty}$ 
domain in $\mathcal{L}^{2}$, a Hilbert space of square integrable functions. 
In this case the evolution of the test quantum wave packet is not uniquely 
determined by the initial wavefunction, the spacetime metric and the 
manifold.

\par An operator, $A$, is called self-adjoint if

\begin{enumerate}

	\item[(i)] $A = A^{\dagger}$

	\item[(ii)] $Dom(A) = Dom(A^{\dagger})$

\end{enumerate}

\noindent where $A^{\dagger}$ is the adjoint of $A$. An operator is 
essentially self-adjoint if (i) is met and (ii) can be met by expanding the 
domain of the operator or its adjoint so that it is true \cite{RS}.

\par A relativistic scalar quantum particle with mass $M$ can be described 
by the positive frequency solution to the Klein-Gordon equation

\begin{equation} \frac{\partial^2\Psi}{\partial{t^2}}=-A\Psi \end{equation}

\noindent in a static spacetime where the spatial operator

\begin{equation} A=-VD^{i}(VD_{i})+V^{2}M^{2} \nonumber \end{equation}

\noindent with $V=-\xi_{\nu}\xi^{\nu}$. Here $\xi^{\nu}$ is the timelike 
Killing field and $D_{i}$ is the spatial covariant derivative on the static slice 
$\Sigma$. The Hilbert space is $\mathcal{L}^{2}(\Sigma)$, the space of 
square integrable functions on $\Sigma$. 

\par If we initially define the domain of $A$ to be $C_{0}^{\infty}(\Sigma)$, 
$A$ is a real, positive, symmetric operator and self-adjoint extensions 
always exist \cite{RS}. If there is only a single, unique extension $A_{E}$, 
then $A$ is essentially self-adjoint. In this case, the Klein-Gordon equation 
for a free scalar particle takes the form \cite{HM}:

\begin{equation} i\frac{d\Psi}{dt}=A_E^{1/2}\Psi \end{equation}

\noindent with

\begin{equation} \Psi(t)=exp(-it(A_E)^{1/2})\Psi(0). \nonumber\\ 
\end{equation}

These equations are ambiguous if $A$ is not essentially self adjoint. This 
fact led Horowitz and Marolf to define quantum mechanically singular 
spacetimes as those in which $A$ is not essentially self-adjoint. Examples 
are considered by Horowitz and Marolf \cite{HM}, Kay and Studer \cite{KS}, 
Helliwell and Konkowski \cite{HK}, Helliwell, Konkowski and Arndt \cite{HKA}, 
Konkowski, Helliwell and Wieland \cite{HKW}, and Konkowski, Reese, Helliwell 
and Wieland \cite{KRHW}.

\par The definition of quantum singularity as originally stated by Horowitz 
and Marolf \cite{HM} applies only to the Klein-Gordon scalar field wave 
operator; however, it is easily extended to Maxwell and Dirac fields 
\cite{HKA}. We say that a spacetime is quantum mechanically singular with 
respect to a Maxwell or Dirac field if the spatial portion of any component of 
the field operator fails to be essentially self-adjoint. We take the Hilbert 
space to be $\mathcal{L}^{2}$ and the original domain to be 
$C_{0}^{\infty}$. To test for essential self-adjointness of the spatial portion 
$A$ of a component of the operator we use the von Neumann \cite{VN} 
criterion. It involves setting $A^{*}\Psi = \pm i \Psi$ and determining the 
number of solutions that belong to $\mathcal{L}^{2}$ for each $i$. If the 
deficiency indices are $(0,0)$, so that no solutions are square integrable, 
then the operator is essentially self-adjoint.

\section{GLT Spacetimes -- Quantum Aspects}

\par In this section we will consider various wave operators in GLT 
spacetime, determine for which modes the operators are essentially 
self-adjoint and show that the GLT spacetimes are generically quantum 
mechanically singular.

\subsection{Scalar Particles}

The Klein-Gordon equation $\square \Phi = M^{2} \Phi$ can be separated in 
GLT spacetime \cite{HKA,HK}. Here

\begin{equation} 	\Phi \sim 
e^{im\phi} e^{ikz} e^{-i\omega t} R(r). \end{equation}

The spatial derivative operator fails to be essentially self-adjoint for $\Phi$ 
modes with

\begin{equation} 	-1 < \frac{m- 
\gamma k}{\beta} < 1 \end{equation}

\noindent where $m$ and $k$ are separation constants, $m$ being the 
azmuthal quantum number and $k$ the momentum.

\subsection{Null Vector Particles}

The classical source-free Maxwell equations $A^{;\nu}_{\mu;\nu}=0$ in the 
Lorentz gauge $A^{\mu}_{;\mu}=0$ can be separated in the GLT spacetime 
by taking linear combinations of modes \cite{HKA}. Here

\begin{equation} 	A_{\mu} \sim 
e^{im\phi} e^{ikz} e^{-i\omega t} R_{\mu}(r). \end{equation}

The spatial derivative operator fails to be essentially self-adjoint for 
$A^{\mu}$ modes with

\begin{equation} 	-1 < \frac{m- 
\gamma k}{\beta} < 1. \end{equation}

\noindent The same as for scalar particles.

\subsection{Free Spin-1/2 Particles}

The Dirac equation $i \gamma^{\alpha}\Psi_{;\alpha} = M\Psi$ for spin-1/2 
particles can be separated in the GLT spacetime \cite{HKA}. Here

\begin{equation} 	\Psi \sim  
{\sqrt{(E+M)} R_{1}(r) \choose i\sqrt{(E+M)} R_{2}(r) e^{i \phi} } e^{-iEt} 
e^{im\phi} e^{ikz}. \end{equation}

The spatial derivative operator is essentially self-adjoint for $\Psi$ modes 
with

\begin{equation} 	-\frac{3}{2} < 
\frac{m-\gamma k +1/2}{\beta} < \frac{3}{2}. \end{equation}

\subsection{Summary}

It is therefore clear that no matter which quantum particle type (scalar, null 
vector or spinor) is used to test the GLT spacetimes for quantum 
mechanical singularity, the generic result is quantum singularity. This is due 
to the fact that specific wave modes are not usually chosen to make the 
spatial wave operator essentially self-adjoint and with general modes the 
operators are not essentially self-adjoint.

\section{Special Cases - Quantum Aspects}

Here we consider special cases of GLT spacetime and test each for quantum 
singularity using a Klein-Gordon field.

\subsection{Minkowski Spacetime}

\par GLT spacetime reduces to Minkowski spacetime if $\beta^{2} = 1$ and 
$\gamma =0$. Both $m=0$ modes (with Bessel function $J_{0} \sim 1$ and 
Neumann function $N_{0} \sim \ln(r)$) are $\mathcal{L}^{2}$, but $r=0$ is a 
regular surface within the spacetime so the $N_{0}$ mode is excluded. 
Therefore the spatial Klein-Gordon wave operator $A$ is essentially self- 
adjoint and the spacetime is quantum mechanically nonsingular. (A well-known 
result presented here for completeness.)

\subsection{Idealized Cosmic String}

\par GLT spacetime reduces to the idealized cosmic string spacetime if 
$\beta^{2} \neq 1$ and $\gamma =0$. Both $m=0$ modes (with Bessel 
function $J_{0} \sim 1$ and Neumann function $N_{0} \sim \ln(r)$) are 
$\mathcal{L}^{2}$, but $r=0$ is NOT a regular surface within the spacetime 
and $N_{0}$ mode cannot be excluded. Therefore the spatial Klein-Gordon 
wave operator $A$ is not essentially self-adjoint and the spacetime is 
quantum mechanically singular. (For details, see \cite{KS, HM, HK, HKA}.)

\subsection{Screw Dislocation Spacetime}

GLT spacetime reduces to the screw dislocation spacetime if $\beta^{2} = 
1$ and $\gamma \neq 0$. There is a continuous infinity of double 
square-integrable modes for each $m$ with $-1 < m - \gamma k < 1$. 
Therefore the spatial Klein-Gordon operator is not essentially self-adjoint and 
the spacetime is quantum mechanically singular. (For details, see \cite{HK}.)

\section{General Cylindrically Symmetric Spacetimes with a Disclination and a 
Dislocation}

\par  A particularly convenient way to establish essential self-adjointness in 
the spatial operator of the Klein-Gordon equation is to use the concepts of 
limit circle and limit point behavior. The approach is as follows.  The Klein- 
Gordon equation for the spacetimes considered in this section can be 
separated in the coordinates $t, \rho, \theta, z$. Only the radial equation is 
non-trivial.  With changes in both dependent and independent variables, the 
radial equation can be written as a one-dimensional Schr\"{o}dinger equation

\begin{equation}     H\Psi(x) = E\Psi(x) \end{equation}

\noindent where $x \in (0,\infty )$ and the operator  $H = - d^{2}/dx^{2} + 
V(x)$.

\par Here we will use this technique to study the general cylindrically 
symmetric static spacetime with a disclination and a dislocation. The metric 
is given by

\begin{equation} 	ds^{2} = e^{-2U} 
[ e^{2K} (d\rho^{2} - dt^{2}) + \rho^{2} B^{2} d\phi^{2}] + e^{2U} [ dz + A 
d\phi]^{2} \end{equation}

\noindent where $U, K, B. A$ are functions of $\rho$ alone. (This metric 
form is taken from SKMHH 22.1 and 22.3 with a slight change in notation 
\cite{SKMHH}; if $B^{2}=1$ this metric agrees with the metric Eq. 1.1 in 
Malcolm's 1998 paper \cite{MM}). Here we will further restrict B to be a 
positive constant. The coordinate ranges are the usual.

\par The classical singularity structure depends on $U,K,B,A$ and can be 
determined using the usual tests for each particular case under 
consideration.

\par The quantum singularity structure will be tested using Weyl's limit 
point-limit circle criterion \cite{weyl} and applying applicable theorems taken 
from Reed and Simon \cite{RS}. The Klein-Gordon wave equation $\Box \Phi = 
M^{2}\Phi$ has mode solutions given by

\begin{equation} 	\Phi \sim e^{- 
i\omega t} e^{ikz} e^{im\phi} H(\rho) \end{equation}

\noindent where

\begin{equation} 	H,_{\rho \rho} + 
\frac{1}{\rho} H,_{\rho} + \{ \omega^{2} - M^{2} e^{-2U} e^{2K} - k^{2} 
e^{-4U} e^{2K} - \rho^{-2}e^{2K} B^{-2}(m- kA)^{2}\}H =0. \end{equation}

\noindent Here square integrability is judged by

\begin{equation} \int d\rho \sqrt{\frac{-g_{3}}{g_{00}}} H^{*}H = \int d\rho 
\rho B H^{*}H. \end{equation}

\noindent If we change variables by letting $H=x^{-1/2}\psi$ and 
$x=\sqrt{B} \rho$, then square integrability is judged by $\int \psi^{*}\psi 
dx$ and the radial equation takes one-dimensional Schr\"odinger form of Eq. 
13. Explicitly,

\begin{equation} 	\psi,_{xx} +  (E - 
V(x))\psi  = 0 \end{equation}

\noindent where $E=\omega^{2}/B$ and

\begin{equation} 	V(x) = 
\frac{M^{2}}{B} e^{-2U} e^{2K} + \frac{k^{2}}{B} e^{-4U} e^{2K} + 
\frac{1}{B^{2}x^{2}} e^{2K} (m-kA)^{2} - \frac{1}{4x^{2}}. \end{equation}

\par We can now use the following \footnote{This section is based on {\bf 
Appendix to X.1} in Reed and Simon \cite{RS}}  to study its limit point-limit 
circle behavior and determine the essential self-adjointness of the spatial 
operator:

\newtheorem{Definition}{Definition} \begin{Definition} The potential $V(x)$ is 
in the limit circle case at $x = 0$  if for some, and therefore for all $E$, {\it 
all} solutions of Eq. 18 are square integrable at zero.  If $V(x)$  is not in the 
limit circle case, it is in the limit point case. \end{Definition}

\par  A similar definition pertains to $x=\infty$.  The potential $V(x)$ is in 
the limit circle case at $x=\infty$ if all solutions of Eq. 18 are square 
integrable at infinity; otherwise, $V(x)$ is in the limit point case at infinity.

\par There are of course two linearly independent solutions of the 
Schr\"{o}dinger equation for given $E$.  If $V(x)$ is in the limit circle case at 
zero, both solutions are $\mathcal{L}^{2}$ at zero, so all linear combinations 
are $\mathcal{L}^{2}$ as well.  We would therefore need a boundary 
condition at $x=0$ to establish a unique solution.  If $V(x)$ is in the limit {\it 
point} case, the $\mathcal{L}^{2}$ requirement eliminates one of the 
solutions, leaving a unique solution without the need of establishing a 
boundary condition at $x=0$. This is the whole idea of testing for quantum 
singularities; there is no singularity if the solution is unique, as it is in the 
limit point case.  The critical theorem is due to Weyl \cite{RS, weyl}.

\newtheorem{theorem}{Theorem} \begin{theorem}[The Weyl limit point-limit 
circle criterion.] If $V(x)$ is a continuous real-valued function on $(0, 
\infty)$, then $H = - d^{2}/dx^{2} + V(x)$ is essentially self-adjoint on 
$C_{0}^{\infty}(0, \infty)$ if and only if $V(x)$ is in the limit point case at 
both zero and infinity. \end{theorem}

\par The following theorem can be used to establish the limit circle-limit 
point behavior at infinity \cite{RS}.

\newtheorem{theorem1}[theorem]{Theorem} \begin{theorem1}[Theorem X.8 
of Reed and Simon \cite{RS}.] If $V(x)$ is continuous and real-valued on $(0, 
\infty)$, then $V(x)$ is in the limit point case at infinity if there exists a 
{\em positive} differentiable function $M(x)$ so that

\begin{enumerate} \item[(i)] $V(x) \ge - M(x)$ \item[(ii)] $\int_{1}^{\infty} 
[M(x)]^{-1/2} dx = \infty$ \item[(iii)] $M'(x)/M^{3/2}(x)$ is bounded near 
$\infty$. \end{enumerate}

Then $V(x)$ is in the limit point case (complete) at $\infty$. \end{theorem1}

A sufficient choice of the $M(x)$ function for our purposes is the power law 
function $M(x) = c x^{2}$ where $c > 0$. Then {\it (ii)} and {\it (iii)} are 
satisfied, so if $V(x) \ge -c x^{2}$, $V(x)$ is in the limit point case at 
infinity.

\par A theorem useful near zero is the following.

\newtheorem{theorem2}[theorem]{Theorem} \begin{theorem2} [Theorem 
X.10 of Reed and Simon \cite{RS}.] Let $V(x)$ be continuous and {\it 
positive} near zero. If $V(x)\ge\frac{3}{4} x^{-2}$ near zero then $V(x)$ is 
in the limit point case.  If for some $\epsilon > 0$, $V(x)\le(\frac{3}{4}- 
\epsilon)x^{-2}$ near zero, then $V(x)$ is in the limit circle case. 
\end{theorem2}

\par Here we can write our $V(x)$ (Eq. 19) as

\begin{equation} 	V(x) = V_{1}(x) - 
\frac{1}{4x^{2}}. \end{equation}

\noindent Then near zero we have the following results:

\begin{itemize}

\item If $V_{1}(x) < \frac{1}{4x^{2}}$, then the theorem does not apply.

\item If $V_{1}(x) \ge x^{-2}$, then $V(x)$ is in the limit point case at $0$.

\item If $\frac{1}{4x^{2}} \le V_{1}(x) \le \frac{(1-\epsilon)}{x^{2}}$ for 
some $\epsilon > 0$, then $V(x)$ is in the limit circle case at $0$.

\end{itemize}

\par Usually, however, it is easiest just to solve the Schr\"odinger equation 
near zero and test the resulting approximate solutions for square 
integrability.

\section{Generalized Levi-Civita Spacetimes with Dislocations}

\par Here we will consider a Levi-Civita (LC) metric that has been generalized 
with the addition of a timelike dislocation ($\alpha \neq 0$) and a spacelike 
dislocation ($\gamma \neq 0$),

\begin{equation} 	ds^{2} = - 
r^{4\sigma} (dt + \alpha d\theta)^{2} + r^{8\sigma^{2} - 4\sigma} dr^{2} + 
r^{8\sigma^{2} -4\sigma} (dz + \gamma d\theta)^{2} + \frac{r^{2- 
4\sigma}}{C^{2}}d\theta^{2}. \end{equation}

\noindent Here $\sigma$ and $C$ are the usual Levi-Civita parameters and 
the coordinate ranges are the usual ones. The constant $\sigma$ is related 
to the mass per unit length of the infinite line mass that the Levi-Civita 
metric can describe, whereas the constant $C^{2} \neq 1$ represents a 
disclination in the spacetime. For a fuller discussion of Levi-Civita 
spacetimes see, for example, Bonnor \cite{bonnor}, Konkowski, Helliwell and 
Wieland \cite{HKW}, and the papers by Herrera et al \cite{HRS, HSTW}. The 
generalized Levi-Civita metric is static if $\alpha = 0$, it reduces to the 
ordinary Levi-Civita metric if $\alpha =0$ and $\gamma = 0$ (see 22.7 of 
SKMHH \cite{SKMHH}), and it reduces to the GLT metric if $\alpha \neq 0,\/ 
\gamma \neq 0,\/ C^{2} \neq 1$ and $\sigma = 0$.

\par The analysis here will be restricted to the static case:

\begin{equation} 	ds^{2} = - 
r^{4\sigma} dt^{2} + r^{8\sigma^{2} -4\sigma} dr^{2} + r^{8\sigma^{2} - 
4\sigma} (dz + \gamma d\theta)^{2} + \frac{r^{2- 
4\sigma}}{C^{2}}d\theta^{2}. \end{equation}

\noindent The classical singularity structure depends on the parameter 
values:

\begin{itemize}

\item $\sigma \neq 0, 1/2$ -- scalar curvature singularity,

\item $\sigma = 0,\/ \gamma =0,\/ C^{2} = 1$ -- Minkowski spacetime -- 
non-singular, of course,

\item $\sigma = 0,\/ \gamma = 0,\/ C^{2} \neq 1$ -- Idealized cosmic 
string -- quasiregular, disclination singularity,

\item $\sigma =0,\/ \gamma \neq 0,\/ C^{2} = 1$ -- Screw dislocation 
spacetime -- quasiregular, dislocation singularity,

\item $\sigma = 1/2$ -- Minkowski spacetime in accelerated coordinates -- 
non-singular.

\end{itemize}

\noindent What about the quantum singularity structure? That too depends 
on the parameter values. We will consider two distinct cases.

\subsection{Generalized LC spacetimes with $\sigma = 1/2$}

The first case we will consider is the $\sigma = 1/2$ case which is Minkowski 
spacetime in accelerated coordinates. The Klein-Gordon equation is separable 
and the radial equation can be written in Schr\"odinger form,

\begin{equation} 	\psi,_{xx} +  (E - 
V(x))\psi  = 0, \end{equation}

\noindent where $E=C^{2}\omega^{2}$,

\begin{equation} 	V(x) = 
C^{2}(k^{2} + M^{2} + m^{2}C^{2})\exp(2 C x), \end{equation}

\noindent and $x = \frac{1}{C}\ln(r)$ with $x \in (-\infty, \infty)$. As $x 
\rightarrow \pm \infty$, $V(x) > -cx^{2}$, so the potential is limit point at  
$\pm \infty$. Therefore, Minkowski spacetime in accelerated coordinates is 
clearly and unambiguously quantum mechanically non-singular.

\subsection{Generalized LC spacetimes with $ \sigma \neq 1/2$}

All other cases can be considered together. Again the Klein-Gordon equation 
is separable and the radial equation can be written in Schr\"odinger form,

\begin{equation} 	\psi,_{xx} +  (E - 
V(x))\psi  = 0, \end{equation}

\noindent where $E = C \omega^{2}/\beta, \beta = (2\sigma - 1)^{2}$,

\begin{eqnarray} 	V(x) = 
(Ck^{2}/\beta)(\beta C x^{2})^{(-\beta +1)/\beta} + (CM^{2}/\beta)(\beta 
C x^{2})^{2\sigma/\beta} \nonumber \\ + (m- \gamma k)^{2} 
\frac{C^{2}}{\beta}(\beta C x^{2})^{(4\sigma - 1)/\beta} - \frac{1}{4 
x^{2}} \end{eqnarray}

\noindent with

\begin{equation} 	x = \frac{1}{C} 
\frac{r^{(2\sigma - 1)^{2}}}{2\sigma - 1} \end{equation}

\noindent for $x \in (0, \infty)$.

\par As $x \rightarrow \infty$, $V(x) > -cx^{2}$, so the potential is limit 
point at infinity. And, as $x \rightarrow 0$,

\begin{itemize}

\item {\bf $\sigma \neq 0$:} $V(x) \rightarrow -\frac{1}{4x^2}$. The 
asymptotic forms of the two independent solutions to the Schr\"odinger 
equation are $\psi_{1} \sim x^{1/2}$ and $\psi_{2} \sim x^{1/2} \ln(x)$. 
Both are $\mathcal{L}^{2}$ so the potential $V(x)$ is limit circle at zero. 

\item {\bf $\sigma = 0$:} $V(x) \rightarrow -\frac{1/4 -(m-\gamma k)^{2} 
C^{2}}{x^2}$. The asymptotic forms of the two independent solutions to the 
Schr\"odinger equation are $\psi_{1} \sim x^{1/2 + |m - \gamma k|C}$, and 
$\psi_{2} \sim x^{1/2} \ln(x)$ if $m = \gamma k$ or $\psi_{1} \sim x^{1/2 
- |m - \gamma k|C}$ if $m \neq \gamma k$. Therefore, $V(x)$ is limit circle 
at zero if $|m - \gamma k|C < 1$ (except the special case $\gamma = 0$ 
Minkowski spacetime where the irregular $\psi_{2}$ solution is discarded at 
zero because $x=0$ is a regular hypersurface in the spacetime), and $V(x)$ 
is limit point at zero if $|m - \gamma k|C \ge 1$.

\end{itemize}

\subsection{Results}

The following results were obtained:

\begin{itemize}

\item $\sigma = 0,\/C^{2} =1,\/ \gamma = 0$: Minkowski spacetime. Here 
$x = 0$ is a regular hypersurface in the spacetime so the $\psi_{2}$ modes 
are discarded and the potential is limit point. Minkowski spacetime is 
quantum mechanically non-singular (a well-known result repeated for 
completeness).

\item $\sigma = 1/2$: Minkowski spacetime in accelerated coordinates. The 
potential $V(x)$ is limit point. Minkowski spacetime in accelerated 
coordinates is quantum mechanically nonsingular.

\item $\sigma = 0$ ($C^{2} \neq 1$ and/or $\gamma \neq 0$) and $\sigma 
\neq 0$: The potential $V(x)$ is limit circle. These generalized LC spacetimes 
are quantum mechanically singular.

\end{itemize}

\noindent These agree when $\gamma =0$ with the results obtained by 
Konkowski, Helliwell and Wieland \cite{HKW} for ordinary LC spacetimes.

\section{Chitre et al Spacetimes}

Under consideration next are a family of spacetimes discovered by Chitre et 
al \cite{Chitre}. Their metric is

\begin{equation} 	ds^{2} = \rho^{- 
4/9} \exp(a^{2} \rho^{2/3})(d\rho^{2} - dt^{2}) + \rho^{4/3}d\phi^{2} + 
\rho^{2/3}(dz + a \rho^{2/3}d\phi)^{2}. \end{equation}

\noindent They are described in SKMHH 22.12 \cite{SKMHH}. Here $a$ is a 
constant. The coordinate ranges are the usual ones.

\par The Chitre et al spacetimes are classically singular with a scalar 
curvature singularity at $\rho =0$.

\par Are they quantum mechanically singular? The Klein-Gordon equation is 
separable and the radial wave equation can be written in one-dimensional 
Schr\"odinger form,

\begin{equation} \psi,_{xx} +  (E - V(x))\psi  = 0, \end{equation}

\noindent with $\rho = x$, $E = \omega^{2}$, and

\begin{eqnarray} 	V(x) = M^{2} x^{- 
4/9} e^{a^2 x^{2/3}} + k^{2} x^{-10/9} e^{a^{2}x^{2/3}}\nonumber  \\ + 
x^{-16/9} e^{a^{2}x^{2/3}} (m-kax^{2/3})^{2} - \frac{1}{4x^{2}}. 
\end{eqnarray}

\noindent As $x \rightarrow \infty$, $V(x) > -cx^{2}$, so the potential is 
limit point at infinity. And, as $x \rightarrow 0$, $V(x) \rightarrow - 
\frac{1}{4x^2}$. The asymptotic forms of the two independent solutions to 
the Schrodinger equation are $\psi_{1} \sim x^{1/2}$ and $\psi_{2} \sim 
x^{1/2} \ln(x)$. Both are $\mathcal{L}^{2}$ so the potential $V(x)$ is limit 
circle at zero. Thus, for all $a$ values, the Chitre et al spacetimes are 
quantum mechanically singular.

\section{Melvin Universes}

\par Next we look at Melvin Spacetimes \cite{Melvin} which are given in 22.13 
of the Exact Solutions Book \cite{SKMHH}. This is a one-parameter family of 
spacetimes with metric,

\begin{equation} ds^{2} = - \alpha ^{2} (1 + R^{2})^{2} (dt^{2} -dR^{2}) + 
\frac{\alpha^{2} R^{2}}{(1 + R^{2})^{2}} d\theta^{2} + (1 + R^{2})^{2} 
dz^{2}, \end{equation}

\noindent where $\alpha$ is a constant and the coordinate ranges are the 
usual ones.

\par The Melvin spacetimes are classically non-singular for all values of 
$\alpha$. They are also quantum mechanically non-singular. This is easily 
seen by writing the radial portion of the Klein-Gordon equation in 
Schr\"odinger form,

\begin{equation} 	\psi,_{xx} +  (E - 
V(x))\psi  = 0. \end{equation}

\noindent Here $E = \omega^{2}$, $R = x$, and

\begin{equation} 	V(x) = \alpha 
^{2} k^{2} + M^{2} + \frac{(m^{2} -1/4)}{x^{2}}. \end{equation}

\noindent As $x \rightarrow \infty$, $V(x) > -cx^{2}$, so the potential is 
limit point at infinity. And, as $x \rightarrow 0$, $V(x) \rightarrow 
\frac{(m^{2}- 1/4)}{x^2}$. The asymptotic forms of the two independent 
solutions to the Schr\"odinger equation are $\psi_{1} \sim x^{1/2 + |m|}$, 
and $\psi_{2} \sim x^{1/2} \ln(x)$ if $m=0$ and $\psi_{2} \sim  x^{1/2 - 
|m|}$ if $m \neq 0$. The $\psi_{2} \/(m=0)$ solution is $\mathcal{L}^{2}$ 
but it is not allowed as $x=0$ is a regular hypersurface of the spacetime. 
The potential is thus limit point at zero, also, and the Melvin universes are 
quantum mechanically non-singular for all parameter values.

\section{Generalized Raychaudhuri Spacetimes with a Disclination and 
Dislocations}

\par The last spacetimes under consideration are generalized Raychaudhuri 
spacetimes with a disclination and two dislocations. These are 
generalizations of the ordinary Raychaudhuri spacetimes \cite{Raych} 
described in 22.16 of the Exact Solutions book \cite{SKMHH}. Their metric is

\begin{eqnarray} 	ds^{2} = -a^{2} 
(\ln(b\rho))^{2} (dt + \alpha d\phi)^{2} + a^{2}(\ln(b\rho))^{2} d\rho^{2} + 
a^{2} B^{2} \rho^{2} (\ln(b\rho))^{2} d\phi^{2} \nonumber \\ + a^{-2} 
(\ln(b\rho))^{-2} (dz + A d\phi)^2 \end{eqnarray}

\noindent where $a, b, \alpha, A$, and $B$ are constants and the coordinate 
ranges are the usual ones. If $\alpha$ is not equal to zero there is a timelike 
dislocation, if $A$ is not equal to zero there is a spacelike dislocation, and if 
$B^{2} \neq 1$ there is a disclination. If $\alpha =0$ the spacetimes are 
static. If $\alpha =0, A = 0, B^{2} =1$ then the ordinary two-parameter 
Raychaudhuri spacetimes are recovered with $a$ and $b$ as the only 
parameters.

\par These generalized Raychaudhuri spacetimes all have a scalar curvature 
singularity at $\rho = 0$. What about quantum singularities?

\par Here we will consider only the static case,

\begin{eqnarray} 	ds^{2} = -a^{2} 
(\ln(b\rho))^{2} dt^{2} + a^{2}(\ln(b\rho))^{2} d\rho^{2} + a^{2} B^{2} 
\rho^{2} (\ln(b\rho))^{2} d\phi^{2} \nonumber \\ + a^{-2} (\ln(b\rho))^{-2} 
(dz + A d\phi)^2 \end{eqnarray}

\noindent which has a disclination and spacelike dislocation. For simplicity, 
assume $B$ is positive in the following analysis. The Klein-Gordon equation is 
separable in the metric coordinates and the radial equation can be put into 
Schr\"odinger form,

\begin{equation} 	\psi,_{xx} +  (E - 
V(x))\psi  = 0, \end{equation}

\noindent where $\rho =x$, $E = \omega^{2}/B^{2}$, and

\begin{eqnarray} 	V(x) = 
\frac{M^{2} a^{2}}{B} (\ln(bx))^{2} + \frac{k^{2}a^{4}}{B} (\ln(bx))^{4} 
\nonumber \\ + \frac{1}{B^{2}x^{2}}(m-kA)^{2} - \frac{1}{4x^{2}}. 
\end{eqnarray}

\noindent As $x \rightarrow \infty$, $V(x) > -cx^{2}$, so the potential is 
limit point at infinity. And, as $x \rightarrow 0$, $V(x) \sim [-\frac{1}{4} + 
\frac{(m-kA)^{2}}{B^{2}}]x^{-2}$ and the two independent asymptotic 
solutions to the Schrodinger equation are $\psi_{1}  \sim x^{1/2 + |m- 
ka|/B}$and $\psi_{2} \sim x^{1/2} \ln(x)$ if $m=kA$ or $\psi_{2} \sim 
x^{1/2 - |m-ka|/B}$ if $m \neq kA$. Therefore, $V(x)$ is limit circle at zero 
if $\frac{|m-kA|}{B} < 1$ and $V(x)$ is limit point at zero if $\frac{|m-kA|}{B} 
\geq 1$. The static generalized Raychaudhuri spacetimes are thus quantum 
mechanically singular for Klein-Gordon modes $-1 < \frac{m-kA}{B} < 1$. If 
generic modes are allowed, all static generalized Raychaudhuri spacetimes 
are generically quantum mechanically singular.

\section{Conclusions}

\par The essential (buried) parameters in the spacetimes considered are not 
evident in a local analysis of the metrics as is done by CLASSI. They are 
evident, however, in a global analysis as one finds when examining the 
spacetimes for classical and quantum singularities. In such analyses there is 
a wealth of information that can be mined from the metric structure.

\par I end with a quote that seems apropos to the buried treasure of globally 
essential parameters. It is from Lewis Carroll's {\it Through the Looking 
Glass},

\begin{quote}

	``I see nobody on the road," says Alice. ``I only wish I had such eyes," 
the King remarked in a fretful tone. ``To be able to see nobody. And at that 
distance, too! Why, it's as much as I can do to see real people, by this light."

\end{quote}.

\section{Acknowledgements}

One of us (DAK) was partially funded by NSF grants PHY-9988607 and PHY- 
0241384 to the U.S. Naval Academy. She also thanks Queen Mary, University 
of London, where some of this work was carried out.

\end{document}